\documentclass[reqno]{amsart}
\usepackage{amssymb,amsthm}             
\numberwithin{equation}{section}        
\usepackage{amssymb,amsthm}             

\usepackage[mathscr]{eucal}

\usepackage[dvips]{hyperref}

\usepackage{graphicx}
\graphicspath{pic/}
\usepackage{psfrag}
\usepackage{subfigure} 

\newcommand{\mnote}[1]{}     

\renewcommand{\Re}{\mathbb R}

\newcommand{\half}{\frac{1}{2}}
\newcommand{\eps}{\epsilon}
\newcommand{\la}{\langle}
\newcommand{\ra}{\rangle}
\newcommand{\tr}{\text{\rm tr}}         

\newcommand{\so}{\mathfrak{so}}         

\newcommand{\ii}{\mathbf{i}}

\newcommand{\Bo}{\mathcal B}
\newcommand{\Sp}{\mathcal S}
\newcommand{\FF}{\mathcal F}

\newcommand{\bU}{\bar U}

\newcommand{\Proj}{{\mathbb P}}

\newcommand{\NN}{{\mathbb N}}
\newcommand{\tNN}{{\widetilde{\mathbb N}}}

\newcommand{\TT}{\mathbb T}

\newcommand{\Xcal}{\mathcal X} 
\newcommand{\Ycal}{\mathcal Y} 
 
\newcommand{\Acal}{\mathcal A} 
\newcommand{\Ocal}{\mathcal O} 

\newcommand{\Lie}{\text{Lie}}

\newcommand{\SO}{\text{SO}}

\newcommand{\gl}{\mathfrak{gl}}
\newcommand{\sym}{\mathfrak{sym}}

\newcommand{\Frob}{\mathcal F}

\newcommand{\Ad}{\text{Ad}}

\theoremstyle{plain}
\newtheorem{thm}{Theorem}[section]
\newtheorem{cor}[thm]{Corollary}

\newtheorem{prop}[thm]{Proposition}
\newtheorem{remark}{Remark}[section]

\title[Static many-body systems]{Static self-gravitating many-body systems in 
Einstein gravity}
\author[L. Andersson]{Lars Andersson${}^*$} \email{larsa@math.miami.edu}
\thanks{${}^*$ Supported in part by the NSF, under
contracts no. DMS 0407732 and DMS-0707306.}
\address{Albert Einstein Institute, Am M\"uhlenberg 1, D-14476 Potsdam,
  Germany \and
Department of Mathematics, University of Miami, Coral Gables, FL
33124, USA}

\author[B. Schmidt]{Bernd G. Schmidt} \email{bernd@aei.mpg.de}
\address{Max-Planck-Institut f\"ur Gravitationsphysik,
Albert-Einstein-Institut, Am M\"uhlenberg 1, D-14476 Golm, Germany}

\begin{document}

\date{May 8, 2009}


\maketitle

\tableofcontents

\section{Introduction} 
This paper continues the study of static self-gravitating elastic bodies in
Einstein gravity which was
initiated 
in \cite{ABS}. In that paper, connected bodies were considered. 
Here, we shall consider the problem of constructing static,
elastic, 
many-body systems in Einstein gravity. 

In Newtonian gravity, one may easily construct 
static many-body systems consisting of rigid bodies. Newton showed that the 
potential inside a spherical, homogenous shell of matter is constant. It
follows that a body placed at rest anywhere inside the
shell, will remain at rest. This example generalizes to hollow ellipsoids,
cf. \cite{chandra:ellipsoidal}.

The problem of constructing static, self-gravitating many-body
configurations consisting of elastic bodies in Newtonian gravity was studied 
in
\cite{beig:schmidt:celest}. One of the results proved there is that 
under certain conditions
it is possible to place a small elastic body at a stationary point of the
Newtonian potential of another body. We remark here that the above mentioned
example of a body at rest inside a spherical shell is only possible in the
case of elastic bodies if the smaller body is spherically symmetric and
placed at the center of the spherical shell, see section \ref{sec:sphere}
below. 

The two main steps in the construction used in \cite{ABS}
are the application of the implicit function theorem to construct solutions
of a certain projected version of the reduced, coupled system of Einstein and
elastic equations, and an 
equilibration argument which showed that the solutions thus constructed are
solutions of the full system. The implicit function theorem is used to deform
from a relaxed body without gravity, i.e. with Newton's constant $G=0$, to a
solution of the above mentioned projected, reduced system with $G \ne 0$. 
As was observed in \cite{BS:CQG2005} 
the linearization of the elastic system at a relaxed
configuration has a cokernel corresponding precisely to the Killing fields
of a Euclidean metric.
It is in order to cancel this degeneracy
that one is led to applying a projection to the system. 

In generalizing the method used in the case of a connected body to the case
of multiple bodies, it turns out that the linearized elastic system now has a
cokernel corresponding to the product of the cokernels corresponding to each
body. However, the equilibration argument developed in \cite{ABS}, 
which is essential to
show that the solution constructed is a solution of the full coupled system of
Einstein and elasticity equations, requires that one has a solution
corresponding to a system which is projected along only one of the factors of
this cokernel. 

In order to achieve this situation one must in the case of multiple bodies, 
after solving the system resulting from projecting out the cokernel, perform
a partial equilibration
making use of the additional degrees of freedom one has in a
multi-body system provided by the relative position and orientation of the
component bodies. This step requires certain non-degeneracy conditions on the Newtonian
potential of the bodies, which are completely analogous to
the ones considered in \cite{beig:schmidt:celest}. 

After the partial equilibration, the 
only remaining degeneracy corresponds to
the cokernel associated with one of the component bodies, and the
equilibration argument from \cite{ABS} can be applied to finish the
argument. 

In sections \ref{sec:setup}--\ref{sec:equilibration}, we present the field
equations for self-gravitating elastic bodies and prove the existence of
solutions to this system, for disconnected bodies, 
given certain non-degeneracy conditions on the
relaxed reference system. In section \ref{sec:examples}, we give a few
concrete examples where this construction applies. 

We remark that for
simplicity, the results
in sections
\ref{sec:setup}--\ref{sec:examples} are stated in terms of self-gravitating
two-body systems. However, the method and results apply with essentially no
changes to the construction of $n$-body systems also for $n > 2$. The point
which needs to be noted here is that instead of partially equilibrating one
of the bodies, cf. section \ref{sec:equilibration}, for the case of a system
of $n$ bodies, it is necessary to partially equilibrate $n-1$ of the bodies,
by a procedure completely analogous to the one presented here. We leave the
details to the reader.

The construction of static solutions of the Einstein equations representing 
disconnected elastic bodies raises the question of the necessary conditions
for the existence of such configurations. In particular, what condition
rules out the existence of two static bodies in
Einstein gravity? 

In Newton's theory of gravity it may seem naively clear that two bodies
can not remain at rest under their mutual gravitational force. 
However, it is important to note
that in order to be able to make such a statement, 
one needs a separation condition. For example, consider a hollow sphere 
and place a small spherical body in the center. 
As is well known, the Newton potential
is constant in the interior of a sphere, and therefore such a configuration
is static.

A
natural condition which rules out the existence of a static two-body system
$\Bo_1 \cup \Bo_2$ 
in Newtonian gravity is given by the existence of a plane which separates the
two bodies $\Bo_1$ and $\Bo_2$. 
In this case it is easy to see that no static solution can exist.  
The self force of $\Bo_1$
vanishes. Consider now the component of the total force generated by 
$\Bo_2$ acting on
$\Bo_1$, which is orthogonal to the separating plane. 
This component is clear non-vanishing and 
hence the bodies must begin to move.  

Now consider the same question in
Einstein's theory of gravity. 
As we show in this paper, there are static
two-body configurations. However, in Einstein gravity,
it is not clear 
how to generalize the separation
condition discussed above from the Newtonian case. In the paper
\cite{beig:schoen}, it was proved that there are no static solutions of
Einstein gravity which admit a separating, complete, totally geodesic
hypersurface $\Sigma$. Here, separating can be taken to mean that $\Sigma$
lies in the complement of the bodies. In particular, this rules out a
static two-body configuration where the bodies are separated by a totally
geodesic hypersurface. 

The problem under discussion, which is 
rather simple in Newton's theory, becomes a challenge in the context of 
Einstein's theory of gravity. One
explanation for this additional difficulty is that in Einstein gravity, we 
loose the concept of force. Furthermore, the self
field is no longer a useful concept because the field equations are nonlinear.
M\"uller zum Hagen wrote a PhD thesis on this problem and has a result for two
separated axially symmetric fluid bodies, cf. \cite{MullerzumHagen}.


\section{General setup and solution of the projected system} 
\label{sec:setup}

We adopt the notations and conventions of \cite{ABS}. We here consider the
situation where the reference body has two connected components, 
$$
\Bo = \Bo_1 \cup \Bo_2 .
$$
The bodies are assumed to be 
disjoint domains in $\Re^3_\Bo$, the extended body, 
and to have free boundaries. In
particular, we consider, as in \cite{ABS}, the zero traction boundary
conditions. 
We consider configurations $f: \Re^3_{\Sp} \to \Re^3_{\Bo}$ and deformations 
$\phi: \Re^3_{\Bo} \to \Re^3_{\Sp}$ as in the case of a connected body. Here
$\Re^3_\Sp$ is the space manifold. 
The
same constitutive relations as in \cite{ABS} are assumed to hold. 

We will assume that $\Bo$ satisfies 
the Newtonian equilibrium condition
\begin{subequations} \label{eq:Newt-equilibrium} 
\begin{align}
\int_{\Bo_\ell} \xi^i  \partial_i v
 &= 0, \quad \ell = 1,2 \label{eq:Gdotint-setup} 
\\
\Delta v &= 4\pi  \mathring{\epsilon} \chi_B 
\end{align} 
\end{subequations} 
for any Killing field $\xi$ on
$\Re^3$, 
see the discussion in section \ref{sec:normalized} for explanation of this
condition. Here $\mathring{\epsilon}$ is the rest mass term in the
relativistic stored energy function, see \cite[section 3.3]{ABS} for
discussion.

The field equations are the same as in the case of a
connected body, namely  in Lagrangian frame
\begin{subequations} \label{eq:Euler}
\begin{align}
\nabla_j (e^U \sigma_i{}^j) &= e^U (n \epsilon - \sigma_l{}^l) \nabla_i U
\quad \text{\rm in } f^{-1}(\Bo), \quad \sigma_i{}^j n_j |_{f^{-1}(\partial
\Bo)}=0 \label{elast} 
\\ \Delta_h U &= 4 \pi G e^U(n \epsilon - \sigma_l{}^l)
\chi_{f^{-1}(\mathcal{B})}\quad \text{\rm in } \Re^3_{\Sp}
\label{potential} \\
G_{ij} &= 8 \pi G ( \Theta_{ij} - e^U
\sigma_{ij}\,\chi_{f^{-1}(\mathcal{B})})\quad \text{\rm in } \Re^3_{\Sp}
\label{metric}
\end{align}
\end{subequations}
and in material frame 
\begin{subequations}\label{eq:material}
\begin{align}
\nabla_A(e^{\bar{U}} \bar{\sigma}_j{}^A) &= e^{\bU} [\bar{\epsilon} -
   \frac{\bar{\sigma}_l{}^l}{\bar{n}}]\,\overline{\partial_i U}\quad
   \text{\rm in } \mathcal{B},\quad \bar{\sigma}_i{}^A
   n_A|_{\partial\mathcal{B}}=0 \label{elastmaterial} \\
\overline{\Delta_h U} &= 4 \pi G e^{\bar{U}}(\bar{n} \bar{\epsilon} -
\bar{\sigma}_l{}^l) \chi_{\mathcal{B}}\quad \text{\rm in } \Re^3_{\Bo}
\label{barU} \\
-\frac{1}{2}\overline{\Delta_h h_{ij}} +
Q_{ij}(\overline{h},\overline{\partial h}) &= 2 \overline{(\nabla_iU)}
\overline{(\nabla_j U)} - 8 \pi G e^{\bar{U}} (\bar{\sigma}_{ij} -
\overline{h_{ij}}\,\bar{\sigma}_l{}^l)\chi_{\mathcal{B}}.
\label{barreduced}
\end{align}
\end{subequations}

\subsection{Analytical setting} \label{sec:analytical}
Let 
$B_1 = W^{2,p}(\Bo) \times W^{2,p}_{\delta} \times E^{2,p}_\delta$, and let
$B_2 = [L^p(\Bo) \times B^{1-1/p,p}(\partial \Bo)] \times L^p_{\delta-2}
\times L^p_{\delta-2}$. Then, $B_1$ is a Banach manifold, and $B_2$ is a
Banach space.

The residuals of equations (\ref{eq:material})
define a map $\FF : \Re \times B_1 \to B_2$, $\FF = \FF(G,
Z)$, where we use $Z = (\phi, \bU, \overline{h_{ij}})$ to denote a general
element of $B_1$. We assume that $\phi$ is a diffeomorphism onto its image.
Thus, $\FF$ has components $\FF =
(\FF_\phi, \FF_U, \FF_h)$, corresponding to the components of $B_2$, given by
\begin{subequations}\label{eq:Fdef}
\begin{align}
\FF_\phi &= \left ( \nabla_A(e^{\bar{U}} \bar{\sigma}_j{}^A) - e^{\bU}
[\bar{\epsilon} - \frac{\bar{\sigma}_l{}^l}{\bar{n}}]\,\overline{\partial_i
U} ,\quad \tr_{\partial\Bo} (\bar{\sigma}_i{}^A) n_A \right ) \\ \FF_U &=
\overline{\Delta_h U} - 4 \pi G e^{\bar{U}}(\bar{n} \bar{\epsilon} -
\bar{\sigma}_l{}^l) \chi_{\Bo} \\ \FF_h &= -\frac{1}{2}\overline{\Delta_h
h_{ij}} + Q_{ij}(\overline{h},\overline{\partial h}) - 2 \overline{\nabla_i
U} \overline{\nabla_j U} + 8 \pi G e^{\bar{U}} (\bar{\sigma}_{ij} -
\overline{h_{ij}}\,\bar{\sigma}_l{}^l)\chi_{\mathcal{B}}
\end{align}
\end{subequations}

Recall, cf. \cite[section 3.2]{ABS}, that in setting up the field
equations in the material frame, an extension $\widehat{\phi}$ of $\phi$ from
$\Bo$ to $\Re_{\Bo}$ is used, cf. \cite[Lemma 4.1]{ABS}. The argument used to 
construct $\widehat{\phi}$ carries over without change to the case of a body
with several connected components. 

The equation to be solved is $\FF(G,Z)=0$.  The material form of the
reference state is given by
$$ Z_0 = (\ii, 0 , \hat \delta_{ij} \circ \ii) \in B_1 .
$$ The map $\FF$ defined by (\ref{eq:Fdef}) is easily verified to satisfy
$\FF(0,Z_0) = 0$ and to map $B_1 \to B_2$ locally near the reference state
$Z_0$. As discussed in \cite{ABS}, the map $\FF$ is smooth.

The Frechet derivative $D_2
\FF(0,Z_0)$ can be represented as the matrix of operators
$$
\begin{pmatrix} D_\phi \FF_\phi & D_U \FF_\phi & D_h \FF_\phi \\
0 & \Delta & 0 \\ 0 & 0 & -\half \Delta \end{pmatrix}
$$ 
(where the entries are evaluated at $(0, Z_0)$). 
In particular, the matrix
is upper triangular, and the diagonal entries are isomorphisms, with the
exception for $D_\phi \FF_\phi(0,Z_0)$ which is Fredholm with nontrivial
kernel and cokernel. 

Let $\phi_\ell$ denote the restriction of $\phi$ to $\Bo_\ell$. Then we have
in a natural way $\FF = (\FF_{\phi_1}, \FF_{\phi_2}, \FF_U, \FF_h)$, and 
$D_\phi \FF_\phi(0,Z_0)$ takes the form
$$
\begin{pmatrix} D_{\phi_1} \FF_{\phi_1} & 0 \\
                              0  & D_{\phi_2} \FF_{\phi_2} \end{pmatrix} 
$$

Let $\delta \bar \sigma_i^{\ A}$ denote any combination of the Frechet
derivatives of $\bar \sigma_i^{\ A}$, 
evaluated at $(0, Z_0)$.
Assuming we
use a coordinate system $X^A$ where $V_{123} = 1$, we have 
the relations
$$ 0 = \int_{\Bo_\ell} \xi^i \partial_A (\delta \bar \sigma_i^{\ A}) -
\int_{\partial\Bo_\ell} \xi^i (\delta \bar \sigma_i^{\ A}) n_A,
\quad \ell = 1,2
$$ 
where $\Bo_\ell$ are the connected components of $\Bo$ and 
$n^A$ is the outward normal.  This can be interpreted as saying that
due to the natural boundary conditions, the linearized elasticity operator,
{\em restricted to each component of $\Bo$} is
automatically equilibrated at the reference configuration $(0, Z_0)$.  
It follows that the cokernel of the operator 
$$ D_\phi \FF_\phi (0, Z_0) : W^{2,p}(\Bo) \to [ L^p(\Bo) \times
B^{1-1/p,p}(\partial \Bo)]
$$ 
consists of the space 
$$
\Ycal_1 \oplus \Ycal_2
$$
where for $\ell = 1,2$, 
$\Ycal_\ell$ is the space 
of Killing fields on 
$\Bo_{\ell}$
considered as a subset of $(\Re^3_{\Bo}, \delta_{\Bo})$, where $\delta_{\Bo} =
\ii^* \hat \delta$ is the Euclidean metric on $\Re^3_{\Bo}$ induced from the
Euclidean metric $\hat \delta$ on $\Re^3_{\Sp}$. The fact that the range and
cokernel of $\FF_\phi$ consists of fields on $\Bo$ is due to the fact that we
defined $\FF$ by passing to the material frame.

Similarly, the kernel of $D_\phi \FF_\phi (0, Z_0)$ 
is the space 
$$
\Xcal_1 \oplus \Xcal_2
$$
where $\Xcal_\ell$ is the space of Killing fields on $\Bo_{\ell}$, considered
as a subset of $(\Re^3_{\Bo}, \delta_{\Bo} )$.

Hence, in view of ellipticity, the operator
$$ D_\phi \FF_\phi (0, Z_0) : W^{2,p}(\Bo) \to [ L^p(\Bo) \times
B^{1-1/p,p}(\partial \Bo)]
$$ 
is Fredholm with the finite dimensional kernel and cokernel discussed above.

\subsection{Projections} \label{sec:proj}
Introduce the projection operators $\Proj_{\Bo_\ell}: B_2 \to B_2$, which acts as the
identity in the second and third components of $B_2$ and is defined in the
first component of $B_2$ as the unique projection along  the 
space of Killing fields  on $(\Bo_{\ell}, \delta_{\Bo})$, 
onto the range of $D_{\phi_\ell} \FF_{\phi_\ell} (0,Z_0)$, which
leaves the boundary data in the first component of $B_2$ unchanged.
If we consider each component
$\Bo_\ell$ separetely, the situation is analogous to the one for the case of
a connected body, so that 
$\Proj_{\Bo_\ell} D_{\phi_\ell} \FF_{\phi_\ell}(0,Z_0)$  is
a surjection. 

We now go back to letting the projection operator act on all components of
$\FF$, and define 
$\Proj_{\Bo_1 \cup \Bo_2} = \Proj_{\Bo_1} \oplus \Proj_{\Bo_2}$ by combining the
projections associated with each component of the body. Then we have that  
$\Proj_{\Bo_1 \cup \Bo_2} D_2 \FF (0,Z_0)$ is a surjection.

The projected system has the property, as in the case of a connected body,
cf. \cite{ABS} that, 
$$
\Proj_{\Bo_1 \cup \Bo_2} D_2 \FF (0,Z_0)
$$ 
is a surjection with finite
dimensional kernel. For this reason, the implicit function theorem can be
applied more or less directly to construct solutions to the first system. 

The data for an Euclidean motion is given by $A = (\alpha^i, \beta_{ij})$, 
where $\alpha^i \in \Re^3$ is a translation vector and $\beta_{ij}$ 
is an orthogonal matrix. 
The motion $A$ acts in $\Re^3_{\Sp}$ by $x \mapsto \beta(x+\alpha)$. 
Denote the group of Euclidean motions
by $\Acal$, and let $I \in \Acal$ be the identity. We shall
consider situations where to leading order $\phi$ maps 
$\Bo_2$ to $A \ii(\Bo_2)$. We implement this by putting
conditions on the the 1-jet of $\phi$ at a point $X_2 \in \Bo_2$.  

\subsection{Solving the projected equation} 
\label{sec:first-projected} 

The following
result is analogous to \cite[Proposition 4.3]{ABS}. However, here we
construct a family of solutions to 
the first projected system with parameters $(G,A)$, where $A \in \Acal$. 
The proof is an application of the implicit function theorem. 
\begin{prop} \label{prop:projected:implicit}
Let $\FF : B_1 \to B_2$ be map defined by (\ref{eq:Fdef}) and let
  $\Proj_{\Bo_1 \cup \Bo_2}$ be
  defined as in section \ref{sec:proj}. Let $X_1 \in \Bo_1$, $X_2 \in
  \Bo_2$  be given
  points, and let $A$ be an Euclidean motion. 
Then, for sufficiently small values
  of Newton's constant $G$, and for $A$ sufficiently close to $I$, 
there is a solution $Z = Z(G,A)$, where $Z = (\phi,
  \bU, \overline{h_{ij}})$, to the reduced, projected equation for
  self-gravitating elastostatics given by
\begin{equation}\label{eq:redproj}
\Proj_{\Bo_1 \cup \Bo_2} \FF(G, Z) = 0, 
\end{equation}
satisfying the conditions
\begin{subequations}\label{eq:jetcond}
\begin{align} 
(\phi - \ii)^i(X_1) &= 0, \quad \delta^C{}_i
\delta_{C[A} \partial_{B]}(\phi - \ii)^i (X_1) = 0 \\
(\phi - A \circ \ii)^i(X_2) &= \alpha^i, \quad \delta^C{}_i
\delta_{C[A} \partial_{B]}(\phi - A \circ \ii)^i (X_2) = \beta_{AB}
\end{align}
\end{subequations} 
In particular, for any $\eps > 0$, there is a $G  > 0$,
such that $Z = Z(G,A)$
satisfies the inequality 
\begin{equation}\label{eq:Gbarsmall}
|| \phi - \ii ||_{W^{2,p}(\Bo_1)} + 
|| \phi - A \circ \ii ||_{W^{2,p}(\Bo_2)}+ ||\overline{h_{ij}} - \delta_{ij}
||_{W^{2,p}_\delta} + ||\bU||_{W^{2,p}_\delta} < \eps .
\end{equation}

\end{prop}
%

\section{Equilibration} \label{sec:equilibration} 
In this section we will make use of the solution to the projected system
(\ref{eq:redproj}) to construct solutions of the full system of Einstein
equations for two static elastic bodies.
Given a solution to the projected system (\ref{eq:redproj}) as in
Proposition \ref{prop:projected:implicit}, our first goal is to construct a
family of solutions to (\ref{eq:redproj}) which are equilibrated on one
component of the body. Once this is done, we are in a situation where we are
able to apply 
the equilibration argument presented in \cite[section 5]{ABS} to construct
solutions to the full system of equations for the self-gravitating elastic
body with two components. 

\subsection{Partial Equilibration} \label{sec:partequi} 
For definiteness we shall focus on $\Bo_2$ and construct a curve $Z = Z(G)$ 
such that 
\begin{equation}\label{eq:xib} 
\int_{\Bo_2} \xi^i b_i = 0
\end{equation} 
for all Killing fields $\xi^i$ on $\Re^3_{\Bo}$, where 
$$
b_i =  \nabla_A(e^{\bar{U}} \bar{\sigma}_j{}^A) - e^{\bU}
[\bar{\epsilon} - \frac{\bar{\sigma}_l{}^l}{\bar{n}}]\,\overline{\partial_i
U}
$$
is the first component of $\FF_\phi$. 
We do this by finding $A = A(G)$, with $A(0)=I$, such that 
$Z(G,A(G))$ solves (\ref{eq:xib}) 
as an equation for $A = A(G)$. 

\subsection{The normalized force} \label{sec:normalized} 
Let $(\xi^i_{(\alpha)})_{\alpha =1}^6$ 
be a basis for the space of Killing fields on
$\Re^3_{\Bo}$. It will be convenient to solve (\ref{eq:xib}) by transforming
to the Eulerian frame, using the change of variables formula as in
\cite[section 5.1]{ABS}.

We define the force map 
$\tNN = (\tNN_{(\alpha)}(G,Z(G,A)))_{\alpha = 1}^6$, 
$\tNN : \Re
\times \Acal \to \Re^6$, by 
\begin{equation}\label{eq:tNN-def}
\tNN_{(\alpha)} (G, A) = 
 \int_{\phi(\Bo_2)} \xi_{(\alpha)}^i \circ \phi^{-1} \left [ 
\nabla_j(e^U \sigma_i{}^j) - e^U
(n \epsilon - \sigma_l{}^l) \nabla_i U \right ] d\mu_h
\end{equation} 
where the right hand side is evaluated at $Z(G,A)$.  
The form of the force map in the material frame is easily found by analogy
with (\ref{eq:xib}). We write this as 
$$
\tNN_{(\alpha)} =  \int_{\Bo_2} \xi^i_{(\alpha)} b_i
$$
We will freely make use of the form of $\tNN$ which is most convenient. 
Since there is a factor $G$ in equation (\ref{potential} we see that
$\tNN(0,A) = 0$. Hence it is convenient to introduce a normalied force map
$\NN$ for $G \ne 0$ by setting 
\begin{equation}\label{eq:NN-def}
\NN = G^{-1} \tNN
\end{equation} 
We define $\NN(0,A)$ by taking the limit as $G \to 0$, which is easily shown
to be well defined, see below. It is natural to view $\tNN$ and $\NN$ as
taking values in the dual of $\Lie(\Acal) \cong \Re^6$. If we denote by $\la
\cdot , \cdot \ra$ the pairing between $\Lie(\Acal)$ and its dual, then we
can write eg.
$$
\la \NN , \xi \ra = G^{-1} \int_{\Bo_2} \xi^i b_i
$$

Introducing a new potential $V$ by $GV = U$, we have 
\begin{subequations}\label{eq:NNspace} 
\begin{align} 
\NN_{(\alpha)} &= \int_{\phi(\Bo_2)} 
\xi_{(\alpha)}^i \circ \phi^{-1} \left [ 
\nabla_j(e^{GV} G^{-1} \sigma_i{}^j) - e^{GV}
(n \epsilon - \sigma_l{}^l) \nabla_i V \right ] d\mu_h
\\
\intertext{where from (\ref{potential}), $V$ and $h$ solve}
\Delta_h V &= 4 \pi  e^{GV}(n \epsilon - \sigma_l{}^l)
\chi_{\phi(\mathcal{B})}\quad \text{\rm in } \Re^3_{\Sp} ,\\
G_{ij} &= 8 \pi G ( G^2 \Theta[V]_{ij} - e^{GV}
\sigma_{ij}\,\chi_{f^{-1}(\mathcal{B})})\quad \text{\rm in } \Re^3_{\Sp}
\end{align} 
\end{subequations} 
\subsection{Newtonian Equilibrium condition} 
In order to evaluate $\NN$ at $(0,I)$ we must consider the limit 
$\lim_{G \to   0} \NN$. 
Calculating this limit is equivalent to calculating the derivative 
$\partial_G \tNN(0,I)$. 
Differentiating the system (\ref{eq:NNspace})
 with respect to $G$, at $G=0$, we find 
\begin{subequations}\label{eq:Newt-first} 
\begin{align}
\NN_{(\alpha)} (0,A) &= \int_{A \ii(\Bo_2)} \xi^i \left  [ 
\partial_j(\delta \sigma)_i{}^j) - 
\mathring{\epsilon} \partial_i V
\right ]  ,  
\label{eq:Gdotint-first} 
\\
\Delta V &= 4\pi \mathring{\epsilon} (\chi_{\ii(\Bo_1)} + \chi_{A \ii(\Bo_2)}) 
\end{align} 
\end{subequations} 
where $\delta  \sigma_i{}^j$ is a collection of derivatives of
$\sigma_i{}^j$ with respect to $\phi$.
The first term in the right hand side of (\ref{eq:Gdotint-first}) 
vanishes identically, cf. the discussion in section \ref{sec:analytical}, see
also \cite[\S 4.2]{ABS}. Therefore we have 
\begin{subequations} \label{eq:Newt} 
\begin{align}
\NN_{(\alpha)} (0,A) &= \int_{A \ii(\Bo_2)} \xi_{(\alpha)}^i  \partial_i V
 , 
\label{eq:Gdotint} 
\\
\Delta V &= 4\pi  \mathring{\epsilon} (\chi_{\ii(\Bo_1)} + \chi_{A \ii(\Bo_2)})
\end{align} 
\end{subequations} 
for Killing fields $\xi_{(\alpha)}$. 
%
%
We shall look for a family of solutions of $\NN(G,A) = 0$ of the form $A =
A(G)$, with $A(0) = I$. For this to be possible, 
it is necessary that the condition 
$\NN(0,I)= 0$ holds. Let $V_\ell$ be the Newtonian potential of the
components $\Bo_\ell$. In view of 
the third axiom of Newton, the 
principle of {\em actio est reactio}, cf. \cite[\S 5]{beig:schmidt:celest},
the self-force of a body vanishes. Applying this to each component and to the
whole body we have 
$$
0 = \int_{\Bo_\ell} \xi^i \partial_i V_\ell, \quad \ell = 1, 2, 
$$ 
and 
\begin{equation}\label{eq:N3}
0 = \int_{\Bo_1} \xi^i \partial_i V_2 + \int_{\Bo_2} \xi^i V_1
\end{equation} 
This implies that $\NN(0,I) = 0$ takes the form 
\begin{equation}\label{eq:newt-equi}
0 = \int_{\Bo_2} \xi^i \partial_i V_1 
\end{equation}
for Killing fields $\xi^i$. We are assuming that (\ref{eq:newt-equi}) holds
for the reference configuration, cf. (\ref{eq:Gdotint-setup}). 
In particular we have 
\begin{subequations} \label{eq:Newtsimp} 
\begin{align}
\NN_{(\alpha)} (0,A) &= \int_{A \ii(\Bo_2)} \xi_{(\alpha)}^i  \partial_i V_1
 , 
\label{eq:Gdotintsimp} 
\\
\Delta V_1 &= 4\pi  \mathring{\epsilon} \chi_{\ii(\Bo_1)} 
\end{align} 
\end{subequations} 
for Killing fields $\xi_{(\alpha)}$.

\subsection{Effect of motions on the normalized force} 
Next we consider the derivative $\partial_A \NN(0,I)$. 
We must consider the $A$ derivative of equation (\ref{eq:Gdotintsimp}) 
at $G=0$. To do this, we must 
consider the effect on $A\ii(\Bo_2)$ of its motion in the potential of
$\ii(\Bo_1)$. 
Let
$A = I + \eps \eta + O(\eps^2)$, where $\eta$ is the
infinitesimal motion with data $(\alpha,\beta)$, i.e. the Killing field 
$\eta^i = \alpha^i + \beta^i{}_j x^j$. 
Then we have 
$A^{-1}  = I - \eps \eta + O(\eps^2)$, so that 
$$
\partial_A (A^{-1} ) \big{|}_{A = I}.\eta = - \eta 
$$
By the change of variables formula
$$
\int_{A \ii(\Bo_2)} \xi^i_{(\alpha)} \partial_i V_1 = \int_{\ii(\Bo_2)}
(\xi^i_{(\alpha)} \partial_i V_1) \circ A^{-1}
$$
Differentiating the integral with respect to $A$ at $I$ in the direction
$\eta$ gives 
\mnote{LA: check the sign of the bracket} 
$$
\int_{\ii(\Bo_2)} [\xi_{(\alpha)},\eta]^i \partial_i V_1 - \xi_{(\alpha)}^i \partial_i
\partial_m V_1 \eta^m
$$
The Lie bracket $[\xi_{(\alpha)}, \eta]$ is again a Killing field, and hence
in view of the fact that, by assumption, each component is equilibrated separately,
cf. equation (\ref{eq:newt-equi}), the first term integrates to zero. 
This leads to 
$$
\partial_A \NN_{(\alpha)}(0,I).\eta = 
- \int_{\ii(\Bo_2)} \xi^i_{(\alpha)} \partial_i \partial_m V_1 \eta^m
$$
We may view $\partial_A \NN(0,I)$ as a linear map $\Re^6 \to \Re^6$. If this
is invertible, we may again apply the implicit function theorem and solve
$\NN = 0$. 
\begin{prop} \label{prop:part-equi}
Assume that the reference body $\Bo \subset \Re^3_{\Bo}$ is in equilibrium in
the sense that (\ref{eq:Gdotint-setup}) holds. Let $Z = Z(G,A)$ be the
solution to $\Proj_{\Bo_1 \cup \Bo_2} \FF = 0$ constructed in Proposition
\ref{prop:projected:implicit}, and let $\NN(G,A)$ be the normalized force map
defined by (\ref{eq:NN-def}). 

Suppose that the derivative $\partial_A \NN(0,I) : \Re^6 \to \Re^6$ is invertible. Then
there is an $\epsilon > 0$ and a smooth map $G \mapsto A(G)$, 
$[0,\eps) \to \Acal$ with 
$$
\NN(G,A(G)) = 0
$$
for $G \in [0,\eps)$. 
\end{prop} 


\subsection{Equilibration} 

We are now in a position to apply the method developed in \cite{ABS} to
complete the construction of solutions to the full system
(\ref{eq:material}), which then also gives a solution to
(\ref{eq:Euler}). If the assumptions of 
Proposition \ref{prop:part-equi} hold, then we may assume without loss of
generality that $\Bo_2$ is equilibrated. Therefore we are in a situation
which is completely analogous to that considered in \cite[\S 5]{ABS}, and a
straightforward application of the methods developed there yields the
following result. 
\begin{thm} \label{thm:main-equi} 
Let $Z(G,A)$ be the solution to the reduced, projected system of equations
for a static, elastic, self-gravitating body  
$$
\Proj_{\Bo_1 \cup \Bo_2} \FF = 0 ,
$$ 
constructed in Proposition \ref{prop:projected:implicit}. 
Assume that the normalized force map satisfies 
$$
\NN(G, A(G)) = 0 .
$$
Then in fact $Z(G,A(G))$ 
solves the full system (\ref{eq:Euler}) of equations for a static, elastic,
self-gravitating body. 
\end{thm} 

The following is an immediate corollary of 
proposition \ref{prop:part-equi} and theorem
\ref{thm:main-equi}.

\begin{cor} \label{cor:main-equi} 
Assume that the reference body $\Bo \subset \Re^3_{\Bo}$ is in equilibrium in
the sense that (\ref{eq:Gdotint-setup}) holds. 
Let $Z = Z(G,A)$ be the
solution to $\Proj_{\Bo_1 \cup \Bo_2} \FF = 0$ constructed in Proposition
\ref{prop:projected:implicit}, and let $\NN(G,A)$ be the normalized force map
defined by (\ref{eq:NN-def}). 

Suppose that the derivative $\partial_A \NN(0,I) : \Re^6 \to \Re^6$ is
invertible. 
Then
there is an $\epsilon > 0$ and a smooth map $G \mapsto A(G)$, 
$[0,\eps) \to \Acal$ such that $Z = Z(G,A(G))$ is a solution to the full 
system (\ref{eq:Euler}) of equations for a static, elastic,
self-gravitating body. 
\end{cor} 

We have now reduced the problem of construcing a static 
self-gravitating two-body system to the question of whether the normalized
force map has the property that $\partial_A \NN(0,I)$ is invertible. This is
clearly determined by the properties of the reference body $\Bo$. 
In section
\ref{sec:examples} below, we consider some particular cases. 

\section{Examples} \label{sec:examples} 

In this section we give some examples of situations where the results
developed in this paper apply. In view of corollary \ref{cor:main-equi}, the
it suffices to that normalized force map of the reference body has invertible
Jacobian. This condition on the reference body 
is precisely equivalent to the condition needed for
the case of static elastic Newtonian two-body systems considered in 
\cite[section 5]{beig:schmidt:celest}. In general, for each example
considered there, we have a corresponding example of a an static elastic
self-gravitating two-body system in Einstein gravity. We shall here present
an independent analysis of these constructions. 

\subsection{Small body} 
Here we consider a situation analogous to the one discussed in
\cite[section 5]{beig:schmidt:celest}.
\mnote{LA: it remains here to discuss how to construct examples of $\Bo_1$
  which have a potential with a non-degenerate stationary point outside the
  body} 
Let $\Bo_1$ be given and let $V_1$ be the Newtonian potential of
$\ii(\Bo_1)$. Assume $V_1$ has a non-degenerate stationary point,
which we may
without loss of generality assume to be at the origin $\Ocal$ 
of the cartesian
coordinate systems on $\Re^3_{\Bo}$ and $\Re^3_{\Sp}$. 
Thus, $\partial_i V_1$ is of the form 
\begin{equation}\label{eq:Vform} 
\partial_i V_1 = B_{ij} x^j + O(|x|^2)
\end{equation} 
We may without loss of generality, after rotating the
coordinate system, assume that $B_{ij}$ is diagonal, $B_{ij} = \sigma_i
\delta_{ij}$. 

We consider a test body $\Bo_2$. We may without loss of generality assume
that $\Bo_2$ has its center of mass at the origin $\Ocal$, i.e. 
$$
\int_{\ii(\Bo_2)} x^i = 0, \quad i=1,2,3
$$
Define 
\begin{equation}\label{eq:Jdef} 
J^{ij} = \int_{\Bo_2} x^i x^j
\end{equation} 
We call $J^{ij}$ as the tensor of inertia of $\Bo_2$, it should
however be noted that the standard usage, cf. 
\cite[section   5.3]{goldstein:mechanics} 
is to define the
inertia tensor as 
$$
\int_{\Bo_2} |x|^2\delta^{ij} - x^i x^j
$$
%
We will now show that there is a homothety $F$ of $\Re^3_{\Sp}$ such that $F
\ii(\Bo_2)$ is equilibrated with respect to the Newtonian potential $V_1$,
i.e. 
\begin{equation}\label{eq:Fequi}
\int_{F \ii(\Bo_2)} \xi^i \partial_i V_1 = 0, \quad \forall \text{
  Killing fields } \xi \text{ of $\Re^3_{\Sp}$.} 
\end{equation} 

A homothetic motion of $\Re^3_{\Sp}$ can be written in the form 
$$
F x = \lambda Q(x+p)
$$
where $\lambda > 0$ is a scale factor, 
$Q$ is a rotation and $p \in \Re^3$ is a translation. 
We first consider homotheties of the form $F(\lambda,p)x = \lambda (x + p)$. Then by the
change of variables formula and (\ref{eq:Vform}), we have for $\xi^i =
\alpha^i$, 
$$
\int_{F \ii(\Bo_2)} \xi^i \partial_i V^i = \lambda^4 \int_{\ii(\Bo_2)} 
\alpha^i B_{ij} (x^j + p^j) + O(\lambda^5)
$$
Thus, defining the normalized force map $\Re_+ \times \Re^3 \to \Re^3$ by 
$$
\NN_i(\lambda,p) = \lambda^{-4} \int_{F(\lambda,p) \ii(\Bo_2)} \partial_i V 
$$ 
we have $\NN(0,0) = 0$
and 
$$
\partial_{p^j} \NN_i(0,0) = |\Bo_2| B_{ij}
$$
Thus, if the matrix $B_{ij}$ is invertible, then we may apply the implicit
function theorem to conclude that for small $\lambda >0$, 
there is a smooth function $p(\lambda)$ satisfying $p(0) = 0$
such that with $F(\lambda, p(\lambda)) \ii(\Bo))$ is equilibrated. In the
following we will consider this case only. 

Next, we consider homotheties of the form $F(\lambda,Q) = \lambda (Qx +
p(\lambda))$. 
After applying the transformation $x \to \lambda(x + p(\lambda))$, $\Bo_2$ is
equilibrated with respect to translational Killing fields. Thus it is 
sufficient to consider 
rotational Killing fields $\xi^i (x)= \beta^i{}_j x^j$ and 
motions of the form 
$$
F(\lambda,Q) = \lambda (Q x + p(\lambda)).
$$
The change of variables
formula and (\ref{eq:Vform}) gives, after taking into account the fact that
$p(\lambda) = O(\lambda)$,  
$$
\int_{F\ii(\Bo_2)} \xi^i \partial_i V = 
\lambda^5 \int \beta^i{}_n Q^n{}j x^j B_{im} Q^m{}_k x^k + O(\lambda^6)
$$
Let $\SO(3)$ be the group of rotations of $\Re^3_{\Sp}$, and 
consider the normalized torque map 
$$
\TT: \Re_+ \times \SO(3) \to \Re^3
$$
defined by 
$$
\TT_{(\alpha)}(\lambda,Q) = \lambda^{-5} \int_{\lambda Q \ii(\Bo_2)}
\xi^i_{(\alpha)} \partial_i V, \alpha = 1,2,3
$$
where $\xi_{(\alpha)}^i(x) = \beta^i_{(\alpha) j} x^j$ is a basis for the Lie
algebra of $\SO(3)$, 
$\so(3) \cong \Re^3$, in particular after raising an index we
have $\beta^{ij}_{(\alpha)} = \beta^{[ij]}_{(\alpha)}$. 

We calculate $\TT(0,Q)$  to be 
$$
\int_{\ii(\Bo_2)} \beta^i{}_n Q^n{}_j x^j B_{ik} 
  Q^k{}_m x^m = \beta^{in} B_{ki} Q^n{}_j J^{jm}
 Q^k{}_m 
$$
where 
$$
Q^n{}_j J^{jm} Q^k{}_m = (Q J Q^t)^{nk} 
$$
expresses the fact that rotating the body by $Q$ induces an orthogonal 
similarity
transformation of the inertia tensor $J$. In particular, there is a $Q_0$
such that $J^0 = Q J Q^t$ is diagonal. Due to the fact that $\beta^{in}$ is
skew, we have 
$$
\beta^{in} B_{ki} J^0_{nk} = \half \beta^{[in]} [B,J^0]_{in} = 0
$$
since by assumption $B_{ki}$ is diagonal. 
Thus, after applying a rotation to
$\Bo_2$, we may without loss of generality assume that $J^{ij}$ is diagonal,
$J^{ij} = \rho^i \delta^{ij}$ so that we have 
$$
\TT(0,I) = 0
$$
We now calculate $\partial_Q \TT(0,I).\mu$ for $\mu \in \so(3)$,
i.e. $\mu = -\mu^t$. We have 
$$
\partial_Q (Q J Q^t) .\mu = [\mu,J]
$$
Let $\gl(3)$ be the space of $3\times 3$ matrices, and for $A \in \gl(3)$, 
let $\Ad_A$ be the linear operator defined by 
$\Ad_A B = [A,B]$. 
\mnote{LA: this is the inner automorphism of the Lie algebra, check the
  standard usage of signs for $\Ad$} 
Recall that the Frobenius inner product on the space of
matrices is $\la A, B\ra_{\Frob} = \tr AB^t$. We can now write 
\mnote{LA: check sign here} 
$$
\partial_Q \TT(0,I) =-  |\Bo_2| \la \beta , \Ad_B \Ad_J \mu \ra_{\Frob}
$$
We observe that 
$\Ad_B$ maps $\so(3) \to \sym(3)$ for $B \in \sym(3)$. If $B$ is diagonal,
which is the case we are considering, then $\Ad_B$ maps 
$\so(3)$ into the 
three dimensional subspace of $\sym(3)$ consisting of 
symmetric matrices with vanishing diagonal elements. 
Further, 
$\Ad : \sym(3) \to \so(3)$. Thus we have $\Ad_B \Ad_J : \so(3) \to
\so(3)$. For $B \in \sym(3)$, then using the cyclic property of the trace,
we have  
$$
\la A, \Ad_B C \ra_{\Frob} = \la \Ad_B A, C \ra_{\Frob},
$$
i.e. $\Ad_B$ is self-adjoint with respect to the Frobenius inner
product. Hence, $\partial_Q \TT(0,I)$ is self-adjoint, and 
$$
\la \partial_Q \TT(0,I).\mu , \beta \ra_{\Frob} = 
- |\Bo_2| \la \Ad_B \beta , \Ad_J \mu \ra_{\Frob}
$$
It follows from this identity that if $\Ad_B$ and $\Ad_J$ have trivial kernel
on $\so(3)$, then $\partial_Q \TT(0,I)$ is invertible. Since $J$ is diagonal,
$J_{ij} = \rho_i \delta_{ij}$, we have 
$$
(\Ad_J \mu)_{ij} = (\rho_i - \rho_j) \mu_{ij}
$$
and hence 
$$
||\Ad_J \mu ||_{\Frob} \geq \min_{i\ne j} |\rho_i - \rho_j| \, ||\mu||_{\Frob}
$$
It follows that if the Hessian $B_{ij}$ and the inertia tensor
$J_{ij}$ of $\Bo_2$ are both diagonal, and such that $B_{ij}$ is invertible
with distinct eigenvalues and $J_{ij}$ has distinct eigenvalues, then
$\partial_Q \TT(0,I)$ is invertible. Recalling that the assumption that
$B_{ij}$ and $J_{ij}$ are diagonal can be imposed without loss of generality,
an application of the implicit function theorem proves the following
proposition. 
\begin{prop} \label{prop:smallbody} 
Assume that the Newtonian potential $V_1$ of $\ii(\Bo_1)$ has a
critical point $\Ocal$
such that the Hessian of $V_1$ at $\Ocal$ is invertible and 
has distinct eigenvalues. Then for any given body $\Bo_2$ such that the
inertia tensor $J$, given by (\ref{eq:Jdef}), has distinct eigenvalues, then
for small $\lambda > 0$, there is a
homothetic motion $F x = \lambda (Qx + p)$ such that $F \ii(\Bo_2)$ is
equilibrated in the Newtonian potential of $\Bo_1$, i.e. equation
(\ref{eq:Fequi}) holds.
\end{prop} 

\begin{remark} The existence of reference configurations satisfying the
  assumptions of proposition \ref{prop:smallbody} was 
shown in \cite[section 5.2]{beig:schmidt:celest}.
\end{remark} 
  
\begin{figure}[!bpt]
\centering
\begin{minipage}{0.45\textwidth}
\centering 
\includegraphics[width=1.5in]{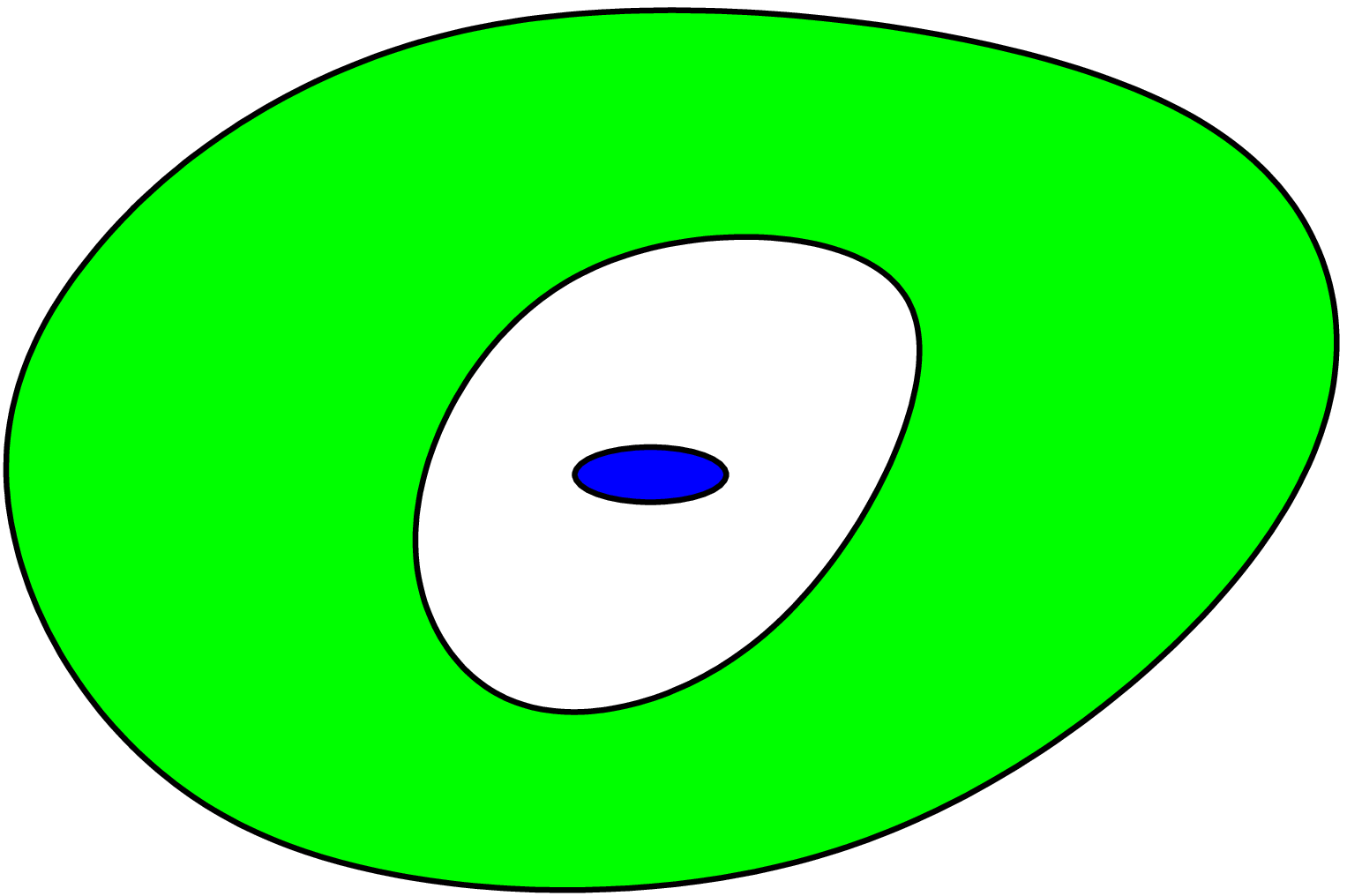}
\end{minipage}%
\begin{minipage}{0.45\textwidth}
\centering 
\includegraphics[width=1.5in]{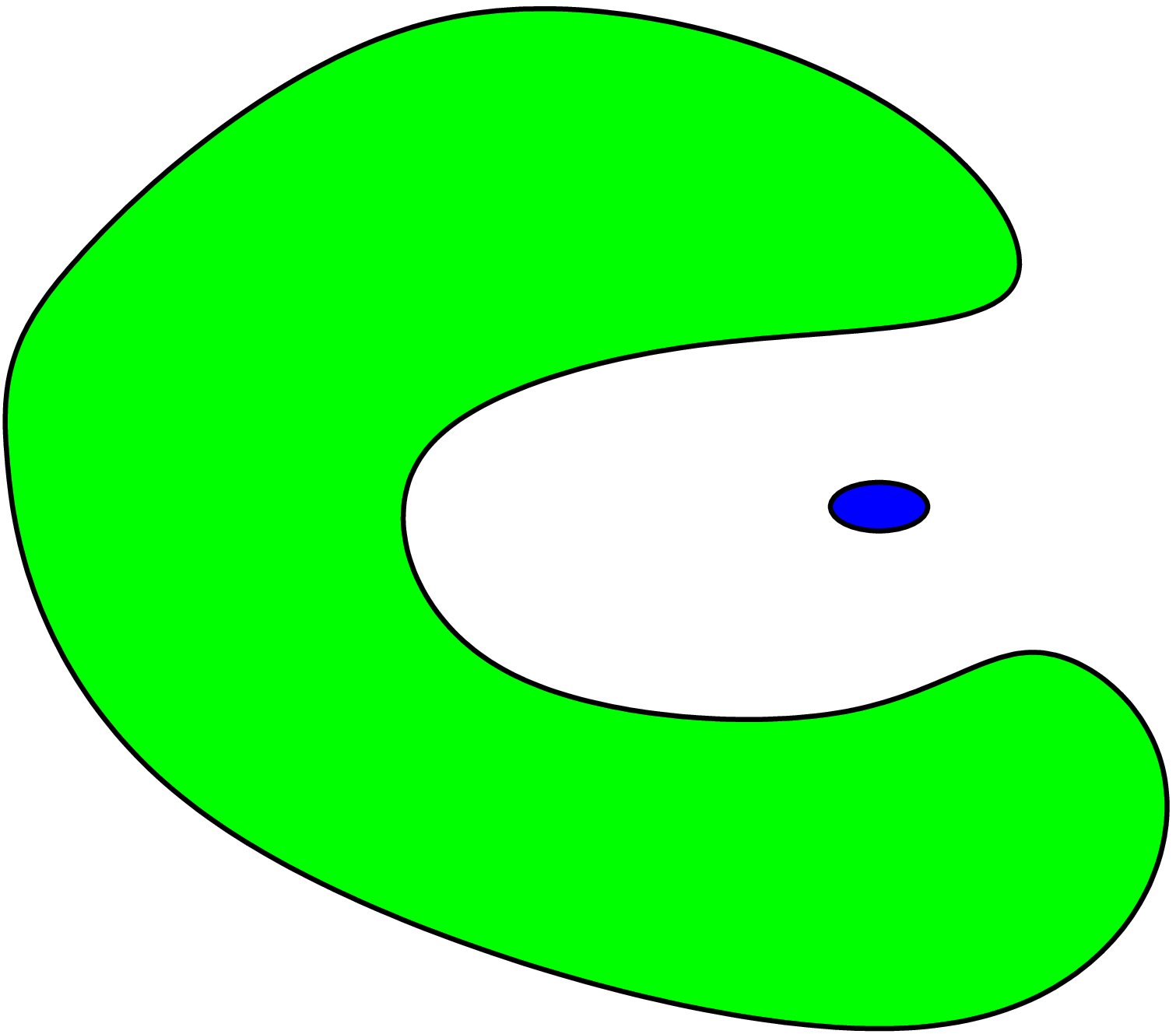}
\end{minipage}%
\caption{Two-body configurations of the type constructed in 
Theorem \ref{thm:smallbody}.}
\label{fig:small}
\end{figure}

We can now apply the results of section \ref{sec:equilibration}, in
particular corollary \ref{cor:main-equi} to deduce the existence of a class
of static,
elastic two-body systems.  Figure \ref{fig:small} illustrated the type of
configurations which are covered by this result. 

\begin{thm} \label{thm:smallbody} 
Assume that the Newtonian potential $V_1$ of $\ii(\Bo_1)$ has a
critical point $\Ocal$, located in the complement of $\ii(\Bo_1)$, 
such that the Hessian of $V_1$ at $\Ocal$ is invertible and 
has distinct eigenvalues, and let $\Bo_2$ be a given body such that the
inertia tensor $J$ of $\Bo_2$ has distinct eigenvalues. 

Then, for
sufficiently small $G$, there is a static, self-gravitating elastic two-body
system in Einstein gravity close to the configuration $\ii(\Bo_1) \cup
F\ii(Bo_2)$, with $F$ a homothetic motion as constructed in proposition
\ref{prop:smallbody}. 
\end{thm}

\subsection{Axisymmetric bodies} 
In this section we consider configurations which have a rotational
symmetry. 
We may without loss of generality assume that $\ii(\Bo_1)$ and $\ii(\Bo_2)$
have have centers of mass at the origin $\Ocal$ 
of the cartesian coordinate system on $\Re^3_{\Sp}$, and that the axis of
symmetry of the bodies is aligned with the $z$ axis. Thus letting $\phi$ be
defined by $\tan \phi = y/x$, we have that $\partial_\phi$ is a symmetry of
the configuration, in the sense that $\partial_\phi \chi_{\Bo_\ell} = 0$,
$\ell = 1,2$.  

Suppose the Newtonian equilibrium condition (\ref{eq:Newt-equilibrium})
holds. 
Let $Z = Z(G,A)$ be the solution to the projected system, constructed using
Proposition \ref{prop:projected:implicit} and let $\NN$ be the normalized
force map as in section \ref{sec:normalized}. For
the present purpose it is convenient to take $\NN$ as defined in terms of the
material frame, i.e. 
$$
\la \NN, \xi \ra = G^{-1} \int_{\Bo_2} \xi^i b_i
$$
As defined, $\NN$ takes values
in the dual of the space of Killing fields of $(\Re^3_{\Sp}, \hat \delta)$.
However, using the Euclidean geometry of
$\Re^3_{\Sp}$, we may consider $\NN$ as taking values in the space of Killing
fields. 
We have $[\NN, \partial_\phi] = 0$. 
One easily checks that the only Killing
fields which have vanishing Lie bracket with $\partial_\phi$ are linear
combindations of $\partial_\phi$ and $\partial_z$. Since $\partial_\phi$ is a
symmetry of the body, it then follows that $\NN$ is proportional to
$\partial_z$. We remark that this can be seen directly from the fact that due
to the axi-symmetry of the body, any load must be along the $z$-axis. 

\begin{figure}[!bpt]
\centering
\begin{minipage}{0.45\textwidth}
\centering
\includegraphics[width=1.2in]{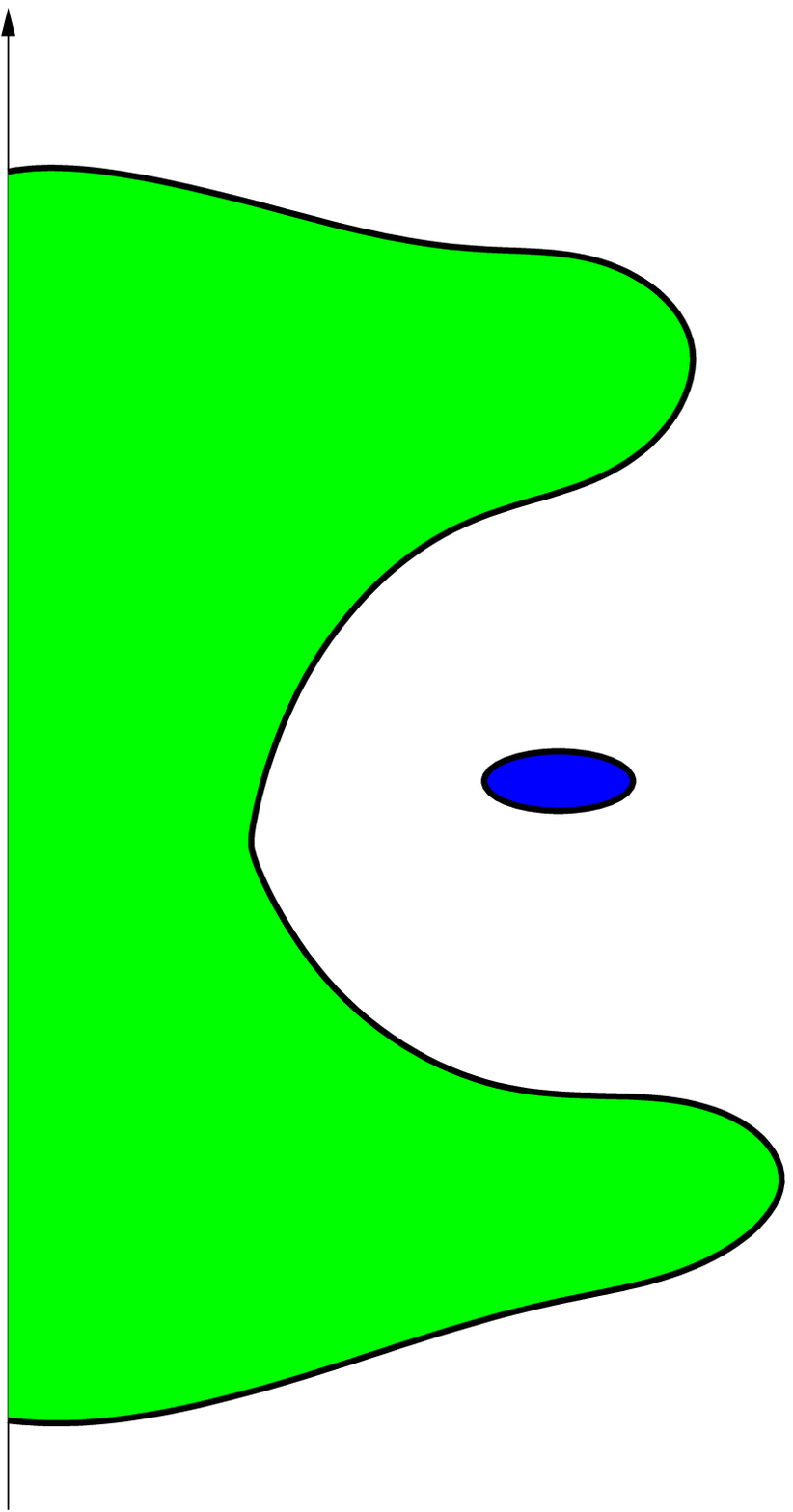}
\end{minipage}%
\begin{minipage}{0.45\textwidth}
\centering 
\includegraphics[width=1.2in]{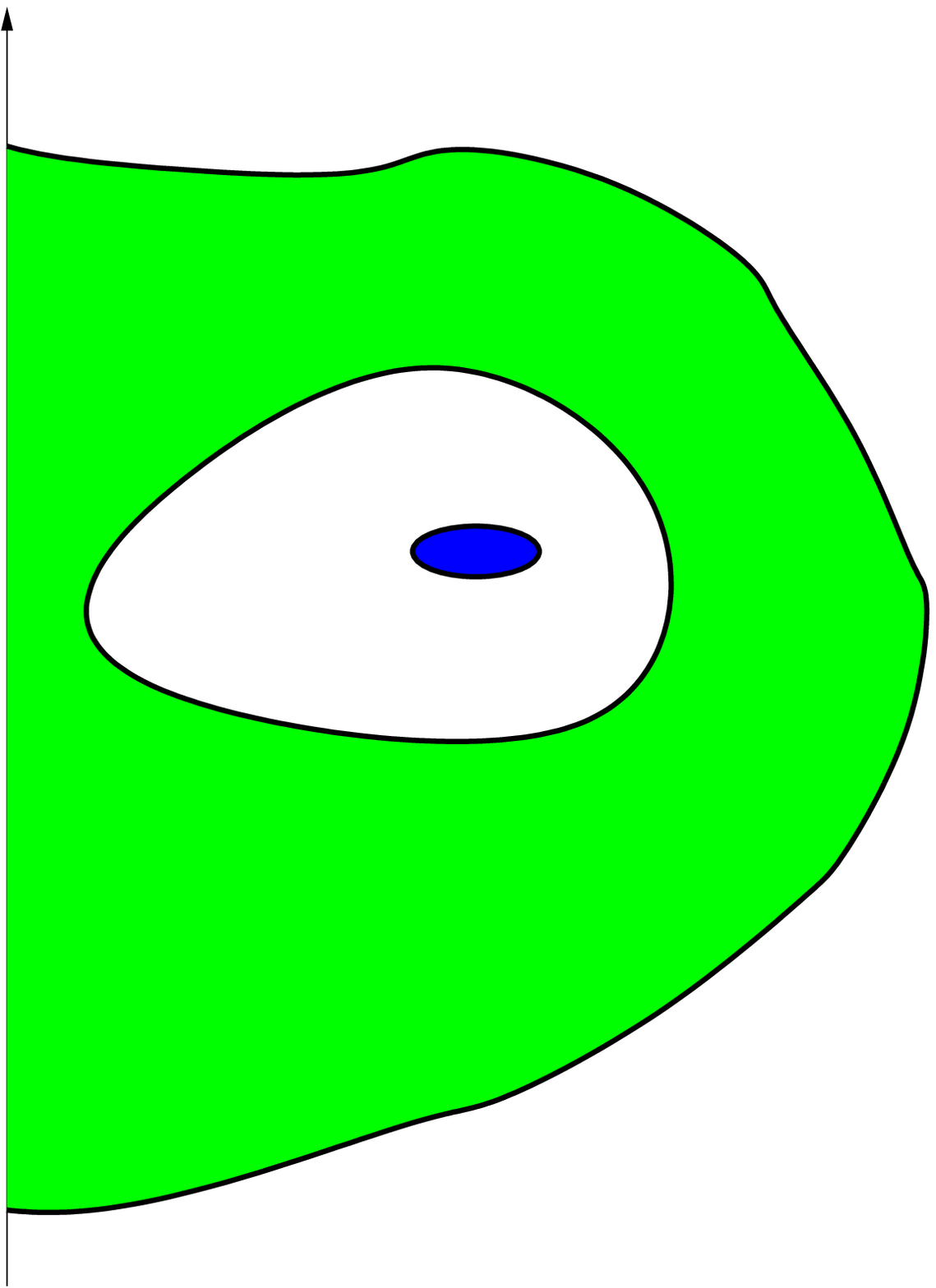}
\end{minipage}%
\caption{Slices through configurations of the type considered in 
theorem \ref{thm:axi}. Each point in the figure corresponds to a circle.}
\label{fig:inside}
\end{figure}

From the above discussion, it follows that the two components cannot be
separated in the $z$-direction. Thus, any axi-symmetric reference
configuration with two components must have the property that one component
is located ``inside'' the other, see figure \ref{fig:inside}. In this figure,
each point corresponds to a circle, i.e. the bodies constructed are achieved
by rotating the regions shown around the $z$-axis. In each case, a  
toroidal object is in equilibrium near, or in a toroidal cavity in 
a larger object. 

Thus, in order to solve the equation $\NN(G,Z(G,A)) = 0$, it suffices to
consider the $\partial_z$ component of $\NN$. Further, the
only motions we need to consider are those of the form $z \to z + a$. 

The terms in the derivative $\partial_A \NN(0,I)$ which we need to consider are
thus of the form 
$$
\int_{\ii(\Bo_2)} \partial_z^2 V_1
$$
If this quantity is non-zero, then in view of the remarks above, the argument
which proves Proposition \ref{prop:part-equi} proves
\begin{prop}\label{prop:axi-equi}
Assume that the reference body $\Bo \subset \Re^3_{\Bo}$ is in equilibrium in
the sense that (\ref{eq:Gdotint-setup}) holds. Let $Z = Z(G,A)$ be the
solution to $\Proj_{\Bo_1 \cup \Bo_2} \FF = 0$ constructed in Proposition
\ref{prop:projected:implicit}, and let $\NN(G,A)$ be the normalized force map
defined by (\ref{eq:NN-def}). 

Suppose $\Bo$ is axisymmetric, with its axis of symmetry aligned with the
$z$-axis, and with center of mass at the origin $\Ocal$. Further, assume that
$\Bo$ is a disjoint union $\Bo = \Bo_1 \cup \Bo_2$ with the property 
that 
$$
\int_{\ii(\Bo_2)} \partial_z^2 V_1 
$$
is nonzero, where $V_1$ is the potential of $\ii(\Bo_1)$. 
Then there is an $\eps > 0$ and a smooth map $G \mapsto A$,
$[0,\eps) \to \Acal$ with $A(G)$ of the form $z \mapsto z+a$ for $a \in \Re$,
  such that 
$$
\NN(G,A(G)) = 0
$$
\end{prop} 
\begin{remark} Proposition \ref{prop:axi-equi} gives an example of a
  situation where the normalized force map $\NN(G,A)$ 
has degenerate derivative at
  $(0,I)$, but where the symmetries of the situation still allow us to 
apply the same argument as in the non-degenerate situation. 
\end{remark} 

\begin{thm} \label{thm:axi} 
Suppose that the reference configuration satisfies the assumptions of
proposition \ref{prop:axi-equi}. Then, for sufficiently small $G$ there is a
static, axisymmetric, self-gravitating two-body system in Einstein gravity,
close to the configuration $\ii(\Bo)$. 
\end{thm}

\subsection{A body in a spherical shell} \label{sec:sphere} 
As mentioned in the introduction, the Newtonian potential inside a spherical
shell consisting of a homogenous material is constant. Considering only rigid
bodies in the Newtonian theory, it is thus possible to place a small body at
rest at an arbitrary position inside the shell. Here we point out that if we
consider instead elastic bodies, then this general construction is no longer
possible. On the contrary, we argue here that the only static configuration
of this type consists of a spherically symmetric body placed at the center of
the shell. 

Consider an outer outer shell
of radius $R$. Take coordinates such that
the center of the ring is at $r=0$. Place the small body at distance $d$
from the center on the $z$--axis and consider  the system in Newtonian
gravity. 

Let us consider linearized elasticity for this system. There 
linearized deformation of the inner body is zero
because the force vanishes in
the interior. Due to the principle of {\em actio est reactio}, 
the force of the inner
body on the outer shell is equilibrated, and hence the
linearized elasticity equation has a solution. 

\begin{figure}[!bpt]
\centering
\psfrag{MAdef}{1:st order deformation} 
\psfrag{MAshell}{shell of radius $R$}
\psfrag{MAF}{$F$} 
\includegraphics[height=3in]{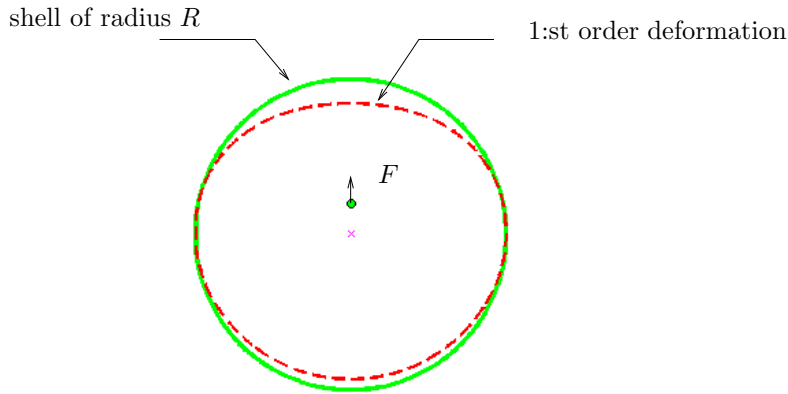}
\vskip -.3truein
\caption{The force generated on a body placed off-center in a spherical,
  elastic shell.}
\label{fig:shell}
\end{figure}

The Newtonian potential generated by the small body is proportional to
$$
\frac{1}{r}+\frac{dz}{r^3}+O(\frac{1}{r^3})
$$
The leading order after the spherical field is a dipole field, i.e. a
$\ell=1$ spherical harmonic. Figure \ref{fig:shell} shows the upper half of the
shell, with the small body. The deformation and the resulting force are
indicated in the figure. 

The linearized deformation generated by $V$ will therefore generate
inside  as dominating contribution a  $l=1$ spherical harmonic
gravitational field
$$
\delta V=az
$$
However, in this field  the small body at $d$ can never be equilibrated
(recall that 
the $l=1$ part is the dominating contribution provided the radius of
the outer shell is suffciently large.) Since the linearized problem has no
solution, we cannot expect a solution to the full non-linear problem to
exist. 

\subsection*{Acknowledgements} LA thanks the Mittag-Leffler Institute, Djursholm, Sweden,
where part of this paper was written, for hospitality and support.


\begin{thebibliography}{1}

\bibitem{ABS}
Lars Andersson, Robert Beig, and Bernd~G. Schmidt, \emph{Static
  self-gravitating elastic bodies in {E}instein gravity}, Comm. Pure Appl.
  Math. \textbf{61} (2008), no.~7, 988--1023.

\bibitem{BS:CQG2005}
Robert Beig and Bernd~G. Schmidt, \emph{Relativistic elastostatics. {I}.
  {B}odies in rigid rotation}, Classical Quantum Gravity \textbf{22} (2005),
  no.~11, 2249--2268.

\bibitem{beig:schmidt:celest}
\bysame, \emph{Celestial mechanics of elastic bodies}, Math. Z. \textbf{258}
  (2008), no.~2, 381--394.

\bibitem{beig:schoen}
Robert Beig and Richard~M. Schoen, \emph{{On Static $n$-body Configurations in
  Relativity}}, Class. Quant. Grav. \textbf{26} (2009), 075014.

\bibitem{chandra:ellipsoidal}
S.~{Chandrasekhar}, \emph{{Ellipsoidal figures of equilibrium}}, New York :
  Dover, 1987., 1987.

\bibitem{goldstein:mechanics}
Herbert Goldstein, \emph{Classical mechanics}, second ed., Addison-Wesley
  Publishing Co., Reading, Mass., 1980, Addison-Wesley Series in Physics.

\bibitem{MullerzumHagen}
M\"uller zum Hagen, PhD thesis.

\end{thebibliography}

\providecommand{\bysame}{\leavevmode\hbox to3em{\hrulefill}\thinspace}
\providecommand{\MR}{\relax\ifhmode\unskip\space\fi MR }
\providecommand{\MRhref}[2]{%
  \href{http://www.ams.org/mathscinet-getitem?mr=#1}{#2}
}
\providecommand{\href}[2]{#2}

\end{document}